\documentclass[epj,twocolumn]{webofc}
\usepackage[varg]{txfonts}   % Web of Conferences font
\woctitle{Flavour changing and conserving processes}
\newcommand{\be}{\begin{equation}}
\newcommand{\ee}{\end{equation}}
\newcommand{\bea}{\begin{eqnarray}}
\newcommand{\eea}{\end{eqnarray}}
\newcommand{\FFa}{{\cal F}_{\pi^0\gamma^*\gamma^*}}
\newcommand{\FFP}{{\cal F}_{{\rm P}\gamma^*\gamma^*}}

\newcommand{\HLbLpi}{\mathrm{HLbL};\pi^0}
\newcommand{\HLbLpione}{\mathrm{HLbL};\pi^0(1)}
\newcommand{\HLbLpitwo}{\mathrm{HLbL};\pi^0(2)}
\newcommand{\HLbLP}{\mathrm{HLbL; P}}

\begin{document}
\title{On the precision of a data-driven estimate of 
the pseudoscalar-pole contribution to hadronic 
light-by-light scattering in the muon $g-2$}

\author{Andreas Nyffeler\inst{1}\fnsep\thanks{\email{nyffeler@kph.uni-mainz.de}}}

\institute{Institut f\"ur Kernphysik and PRISMA Cluster of Excellence,
  Johannes Gutenberg-Universit\"at Mainz, D-55128 Mainz, Germany}

\abstract{%
  The evaluation of the numerically dominant pseudoscalar-pole
  contribution to hadronic light-by-light scattering in the muon $g-2$
  involves the pseudoscalar-photon transition form factor ${\cal
    F}_{{\rm P}\gamma^*\gamma^*}(-Q_1^2, -Q_2^2)$ with ${\rm P} =
  \pi^0, \eta, \eta^\prime$ and, in general, two off-shell photons
  with spacelike momenta $Q_{1,2}^2$. We determine which regions of
  photon momenta give the main contribution for hadronic
  light-by-light scattering. Furthermore, we analyze how the precision
  of future measurements of the single- and double-virtual form factor
  impacts the precision of a data-driven estimate of this contribution
  to hadronic light-by-light scattering.}
%
%%%%%%%%%%%%%%%%%%%%%%%%%%%%%%%%%%%%%%%%%%%%%%%%%%%%%%%%%%%%%%%%%%%%%%%%%%%%
% extra for preprint
\onecolumn
\thispagestyle{empty}
\begin{flushright}
{\large
February 11, 2016 \\[1mm] 
MITP/16-019}
\end{flushright}

\vfill

\begin{center}
{\huge\bf
On the precision of a data-driven estimate of \\[1mm]
the pseudoscalar-pole contribution to hadronic \\[3.5mm]
light-by-light scattering in the muon $g-2$~$^*$}\\[1.5cm]  
{\large Andreas Nyf\/feler}\\[1cm]
{\large Institut f\"ur Kernphysik and PRISMA Cluster of Excellence, \\[1mm]
Johannes Gutenberg-Universit\"at Mainz,   
D-55128 Mainz, Germany}

\vfill

{\large\bf Abstract}\\[3mm]

\begin{minipage}{0.8\textwidth}
  The evaluation of the numerically dominant pseudoscalar-pole
  contribution to hadronic light-by-light scattering in the muon $g-2$
  involves the pseudoscalar-photon transition form factor ${\cal
    F}_{{\rm P}\gamma^*\gamma^*}(-Q_1^2, -Q_2^2)$ with ${\rm P} =
  \pi^0, \eta, \eta^\prime$ and, in general, two off-shell photons
  with spacelike momenta $Q_{1,2}^2$. We determine which regions of
  photon momenta give the main contribution for hadronic
  light-by-light scattering. Furthermore, we analyze how the precision
  of future measurements of the single- and double-virtual form factor
  impacts the precision of a data-driven estimate of this contribution
  to hadronic light-by-light scattering.
\end{minipage}
\end{center}

\vfill
\noindent\rule{8cm}{0.5pt}\\
$^*$ Invited talk at FCCP2015 - Workshop on ``Flavour changing and
conserving processes,'' 10-12 September 2015, Anacapri, Capri Island,
Italy. Some preliminary results have earlier been presented at the Spring
Meeting of the German Physical Society (DPG), Physics of Hadrons and Nuclei,
24 March 2015, Heidelberg, Germany, at PHOTON 2015, 15-19 June 2015,
Novosibirsk, Russia and at the Workshop ``High-precision QCD at low
energy'', 2-22 August 2015, Benasque, Spain. 
\setcounter{page}{0}
\newpage
\twocolumn
% end extra for preprint
%%%%%%%%%%%%%%%%%%%%%%%%%%%%%%%%%%%%%%%%%%%%%%%%%%%%%%%%%%%%%%%%%%%%%%%%%%% 
\maketitle

\section{Introduction}

The anomalous magnetic moment of the muon $a_\mu$ serves as an
important test of the Standard Model (SM)~\cite{JN_09}. Since several
years, there is an intriguing discrepancy of $3-4\sigma$ between the
experimental value~\cite{g-2_exp} and the theoretical SM
prediction~\cite{JN_09, g-2_SM_theory}. While this could be a sign of
New Physics, the hadronic contributions from vacuum polarization (HVP)
and light-by-light scattering (HLbL) have large uncertainties, which
make it difficult to interpret the deviation as a clear sign of
physics beyond the SM. The hadronic uncertainties need to be reduced
and better controlled, also to fully profit from new planned $g-2$
experiments~\cite{Hertzog_talk}. While the HVP
contribution~\cite{HVP_talks} can be improved systematically with
measurements of $\sigma(e^+ e^- \to \mbox{hadrons})$, the often used
estimates for HLbL
\bea
a_\mu^{\mathrm{HLbL}} & = & (105 \pm 26) \times 10^{-11}, \qquad
\mbox{\cite{PdeRV_09}} \\ 
a_\mu^{\mathrm{HLbL}} & = & (116 \pm 40) \times 10^{-11}, \qquad
\mbox{\cite{N_09,JN_09}}  
\eea
are both based on model calculations~\cite{HKS, BPP, KN_02,
  Knecht_et_al_PRL_02, MV_04},\footnote{There are attempts ongoing to
  calculate the HLbL contribution from first principles in Lattice
  QCD. A first, still incomplete, result was obtained
  recently~\cite{Lattice_HLbL_Blum_et_al}. See also the approach
  proposed in~\cite{Lattice_HLbL_Mainz}.} which suffer from
uncontrollable uncertainties, see also \cite{HLbL_talks}. 

In this situation, a dispersive approach to HLbL was proposed recently
in Refs.~\cite{HLbL_DR_Bern_Bonn, HLbL_DR_Mainz}, which tries to
relate the presumably numerically dominant contributions from the
pseudoscalar-poles and the pion-loop to, in principle, measurable form
factors and cross-sections, $\gamma^* \gamma^* \to \pi^0, \eta,
\eta^\prime$ and $\gamma^*\gamma^* \to \pi\pi$, with on-shell
intermediate pseudoscalar states.\footnote{There have been objections
  raised at this meeting~\cite{Vainshtein_talk} about the
  implementation of the dispersive approach for the pion-pole
  contribution. There should be no form factor at the external
  vertex~\cite{MV_04}.}  The hope is that this data-driven estimate
for HLbL will allow a $10\%$ precision for these contributions and
that the remaining, hopefully smaller contributions, e.g.\ from
axial-vectors ($3\pi$-intermediate state) and other heavier states,
can be obtained within models with about $30\%$ uncertainty to reach
an overall, reliable precision goal of $20\%~(\delta
a_\mu^{\mathrm{HLbL}} \approx 20 \times 10^{-11})$.

Most model evaluations of $a_\mu^{\HLbLpi}$ (pion-pole defined in
different ways~\cite{HLbL_DR_Bern_Bonn, HLbL_DR_Mainz, MV_04} or
pion-exchange with off-shell-pion form factors~\cite{N_09, JN_09}) and
$a_\mu^{\HLbLP}$, with ${\rm P} = \pi^0, \eta, \eta^\prime$, agree at
the level of $15\%$, but the full range of estimates (central values)
is much larger:
\bea 
a_{\mu;{\rm models}}^{\HLbLpi} & = & (65 \pm 15) \times 10^{-11} \quad
(\pm 23\%), \label{range_HLbLpi0} \\  
a_{\mu;{\rm models}}^{\HLbLP} & = & (87 \pm 27) \times 10^{-11} \quad
(\pm 31\%). \label{range_HLbLP}
\eea

In this paper we study, within the dispersive approach, which are the
most important momentum regions for the pseudoscalar-pole contribution
$a_\mu^{\HLbLP}$. We also analyze what is the impact of the precision
of current and future measurements of the single- and double-virtual
pseudoscalar transition form factor $\FFP(-Q_1^2, -Q_2^2)$
(TFF)~\cite{TFF} on the uncertainty of a data-driven estimate of this
contribution to HLbL. More details can be found in
Ref.~\cite{Nyffeler_16}.

\section{Pseudoscalar-pole contribution}

\begin{figure}[h!]
\centering 
\hspace*{-0.2cm}\includegraphics[width=8.5cm,clip]{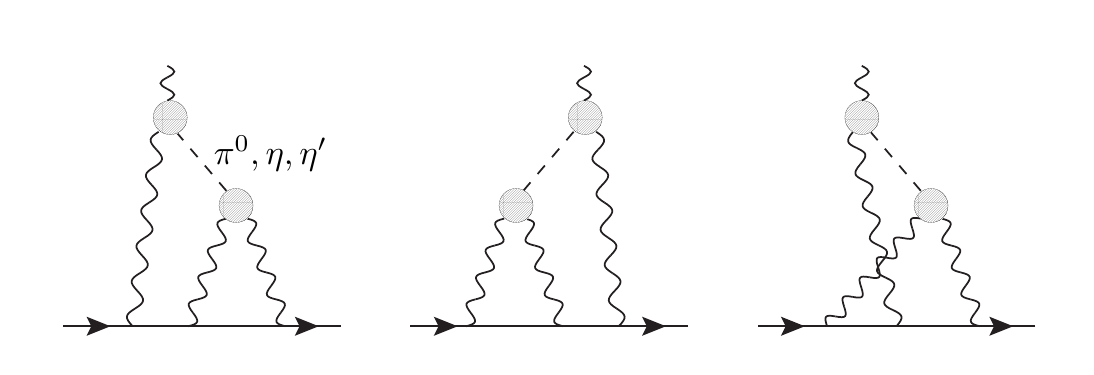}    
\caption{The pseudoscalar-pole contribution to HLbL scattering. The
  shaded blobs represent the transition form factor
  $\FFP(q_1^2,q_2^2)$ where ${\rm P} = \pi^0, \eta, \eta^\prime$.}
\label{Fig:HLbL_PS-poles}
\end{figure}

The dominant contribution to HLbL arises, according to most model
calculations, from the one-particle intermediate states of the light
pseudoscalars $\pi^0, \eta, \eta^\prime$ shown in the Feynman diagrams
in Fig.~\ref{Fig:HLbL_PS-poles}.  We will evaluate only the
pseudoscalar-pole contribution of these two-loop diagrams.  In order
to simplify the notation, we mainly discuss the neutral pion-pole
contribution in this section. The generalization to $\eta$ and
$\eta^\prime$ is straightforward.

The Feynman diagrams for the pion-pole contribution involve the pion
TFF $\FFa(q_1^2,q_2^2)$ which is defined by the QCD vertex function:
\bea  \label{TFF} 
\lefteqn{i \int d^4 x \, e^{i q_1 \cdot x}  \langle 0 | T \{ j_\mu(x)
  j_\nu(0) \}| \pi^0(q_1 + q_2) \rangle} \nonumber \\ 
& & = \varepsilon_{\mu\nu\alpha\beta} \, q_1^\alpha \, q_2^\beta \,  
 \FFa(q_1^2, q_2^2) \, . 
\eea
Here $j_\mu(x) = ({\overline \psi} \hat Q \gamma_\mu \psi)(x)$ is the
light quark part of the electromagnetic current (${\overline \psi}
\equiv ({\overline u}, {\overline d}, {\overline s}$) and $\hat Q =
\mbox{diag}(2,-1,-1)/3$ is the charge matrix). The form factor
describes the interaction of an on-shell neutral pion with two
off-shell photons with four-momenta $q_1$ and $q_2$. It is Bose
symmetric $\FFa(q_1^2,q_2^2) = \FFa(q_2^2,q_1^2)$ and for real photons
it is related to the decay width $\FFa^2(0,0) = 4 \Gamma(\pi^0 \to
\gamma\gamma) / (\pi \alpha^2 m_\pi^3)$. Often the
normalization with the chiral anomaly $\FFa(0,0) = - N_c /(12\pi^2
F_\pi)$ is used. 

The projection of the Feynman diagrams in Fig.~\ref{Fig:HLbL_PS-poles}
on the muon $g-2$ leads to a two-loop integral involving the
propagators of the muon, the photons, the pion, the product of two
transition form factors and some kinematical
functions~\cite{KN_02}. After a Wick rotation to Euclidean momenta and
averaging over the direction of the muon momentum, one can perform for
arbitrary form factors all angular integrations, except one over the
angle $\theta$ between the four-momenta $Q_1$ and $Q_2$ which also
appears through $Q_1 \cdot Q_2$ in the form
factors~\cite{JN_09}. Writing
\be \label{amupi0_start}
a_\mu^{\HLbLpi} =  \left( \frac{\alpha}{\pi} \right)^3 \left[
  a_\mu^{\HLbLpione} + a_\mu^{\HLbLpitwo} \right],  
\ee 
where the first contribution arises from the first two Feynman
diagrams in Fig.~\ref{Fig:HLbL_PS-poles} and the second from the last
graph, one obtains the following three-dimensional integral
representation for the pion-pole contribution~\cite{JN_09}
\bea 
\lefteqn{a_\mu^{\HLbLpione} = \int_0^\infty \!\!\!dQ_1 \int_0^\infty
\!\!\!dQ_2 \int_{-1}^{1} \!\!d\tau \, \, w_1(Q_1,Q_2,\tau)}  \nonumber \\ 
& & \hspace*{-0.4cm}\times \, \FFa(-Q_1^2, -(Q_1 + Q_2)^2) \, 
\FFa(-Q_2^2,0), \label{amupi0_1} \\ 
\lefteqn{a_\mu^{\HLbLpitwo} = \int_0^\infty \!\!\!dQ_1 \int_0^\infty
\!\!\!dQ_2 \int_{-1}^{1} \!\!d\tau \, \, w_2(Q_1,Q_2,\tau)} \nonumber \\  
& & \hspace*{-0.4cm}\times \, \FFa(-Q_1^2, -Q_2^2) \,
\FFa(-(Q_1+Q_2)^2,0). \label{amupi0_2}
\eea 
The integrations in Eqs.~(\ref{amupi0_1}) and (\ref{amupi0_2}) run
over the lengths of the two Euclidean four-momenta $Q_1$ and $Q_2$ and
the angle $\theta$ between them $Q_1 \cdot Q_2 = Q_1 Q_2 \cos\theta$
and where we introduced $\tau = \cos\theta$.\footnote{We have written
  $Q_i \equiv |(Q_i)_{\mu}|, i=1,2$ for the length of the four
  vectors. Following~\cite{HLbL_DR_Bern_Bonn} we changed the notation
  used in ~\cite{JN_09} and write $\tau = \cos\theta$ in order to
  avoid confusion with the Mandelstam variable $t$ in the context of
  the dispersive approach.} 
From the definition of the form factor in
Eq.~(\ref{TFF}) it follows that the weight functions
$w_{1,2}(Q_1,Q_2,\tau)$ are dimensionless. Furthermore
$w_{2}(Q_1,Q_2,\tau)$ is symmetric under $Q_1 \leftrightarrow
Q_2$~\cite{JN_09}. Finally, $w_{1,2}(Q_1,Q_2,\tau) \to 0$ for $Q_{1,2}
\to 0$ and for $\tau \to \pm 1$. Expressions for the precise behavior
of $w_{1,2}(Q_1,Q_2,\tau)$ for $Q_{1,2} \to 0$ and $\tau \to \pm 1$,
as well as for $Q_{1,2} \to \infty$, can be found in
Ref.~\cite{Nyffeler_16}.

The three-dimensional integral representation in Eqs.~(\ref{amupi0_1})
and (\ref{amupi0_2}) separates the generic kinematics in the pion-pole
contribution to HLbL, described by the model-independent weight
functions $w_{1,2}(Q_1,Q_2,\tau)$, from the dependence on the single-
and double-virtual TFF $\FFa(-Q^2,0)$ and $\FFa(-Q_1^2,-Q_2^2)$ in the
spacelike region, which can in principle be measured, obtained from a
dispersion relation (DR)~\cite{pion_TFF_DR, eta_etaprime_TFF_DR} or,
as has been done so far for HLbL, modelled.

The only explicit dependence on the pseudoscalar appears in the weight
functions $w_{1,2}(Q_1, Q_2, \tau)$ via the mass in the pseudoscalar
propagators, i.e.\ a factor $1/(Q_2^2 + m_{\rm P}^2)$ in $w_1$ and a
factor $1/((Q_1 + Q_2)^2 + m_{\rm P}^2) = 1/(Q_1^2 + 2 Q_1 Q_2 \tau +
Q_2^2 + m_{\rm P}^2)$ in $w_2$.  Of course, also the form factors will
depend on the type of the pseudoscalar.

\section{The weight functions $w_{1,2}(Q_1,Q_2,\tau)$}

\begin{figure*}
\centering 
\includegraphics[height=5cm,clip]{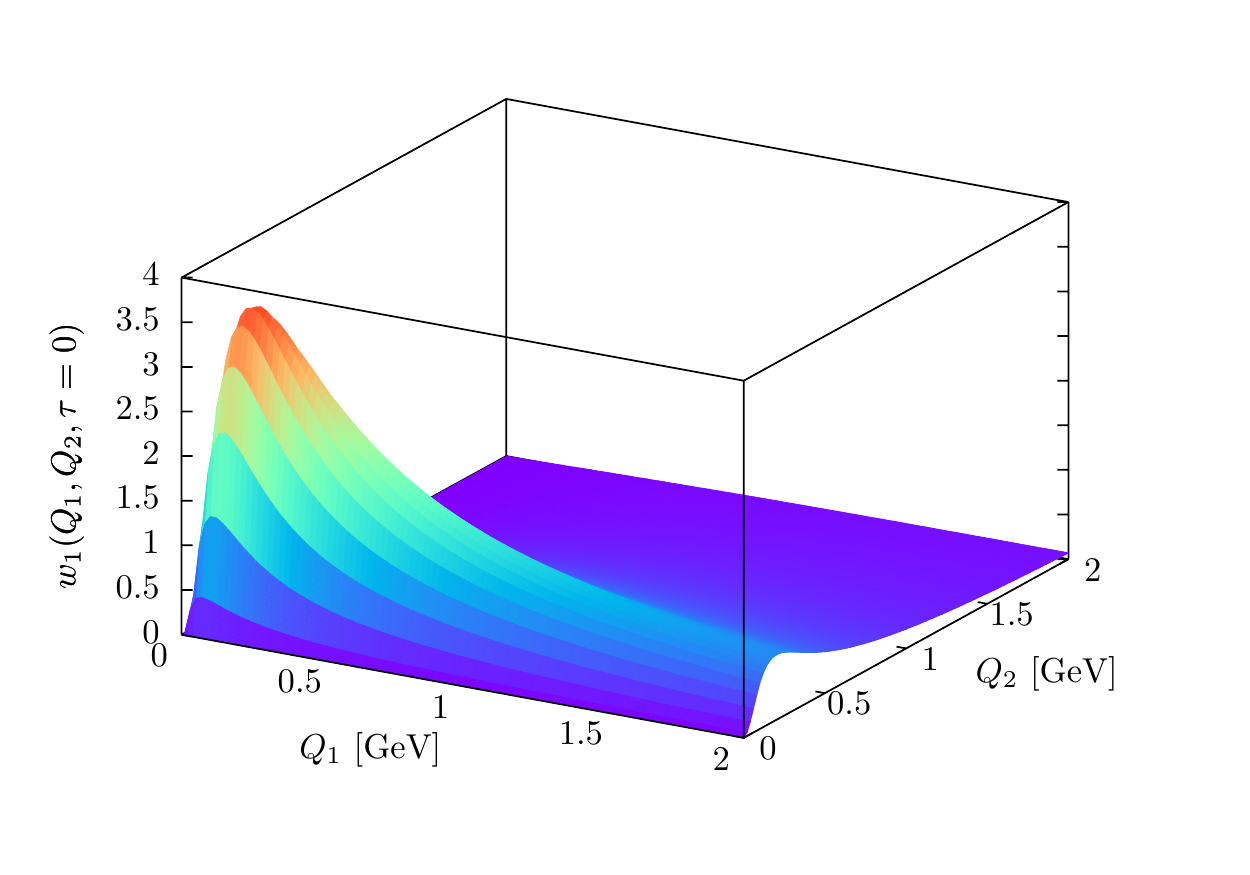}
\hspace*{1cm}\includegraphics[height=5cm,clip]{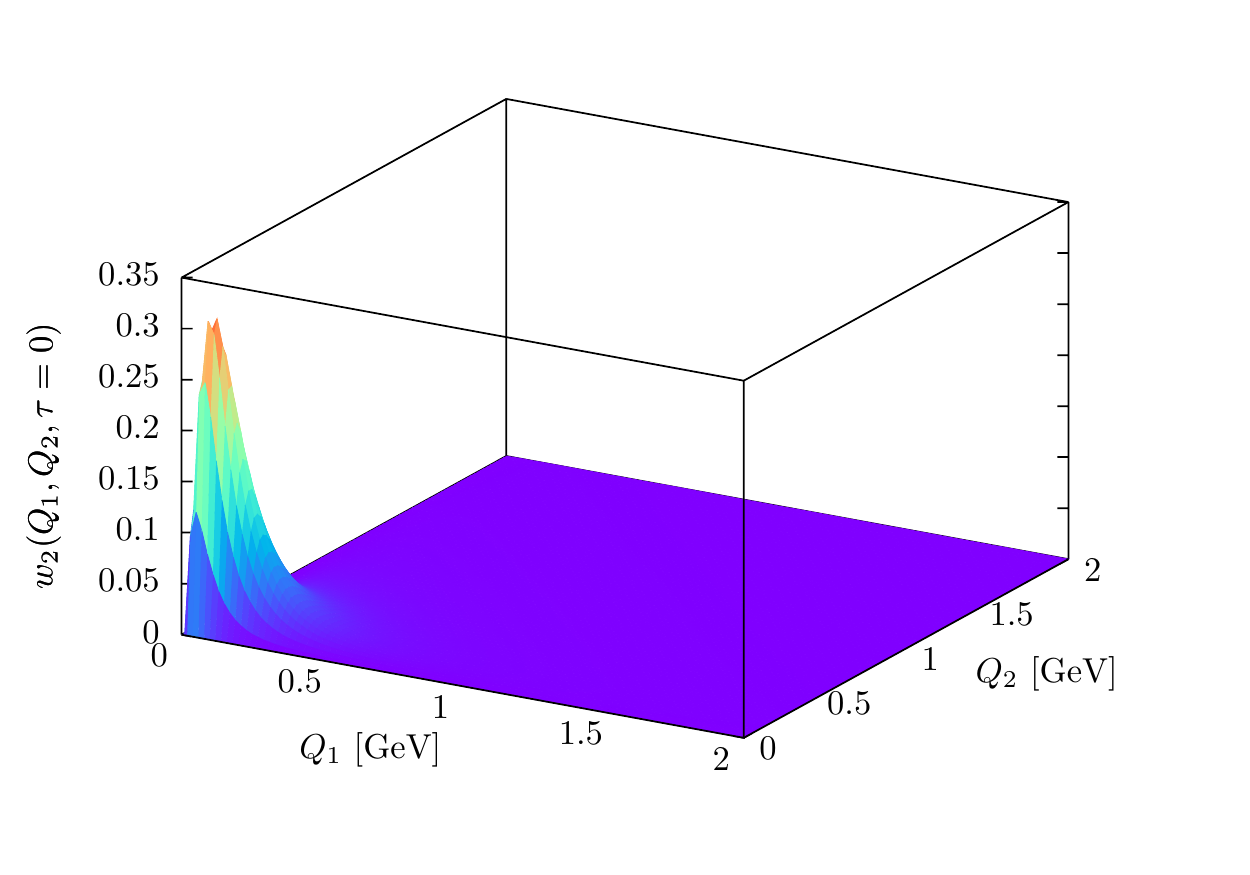}

\includegraphics[height=5cm,clip]{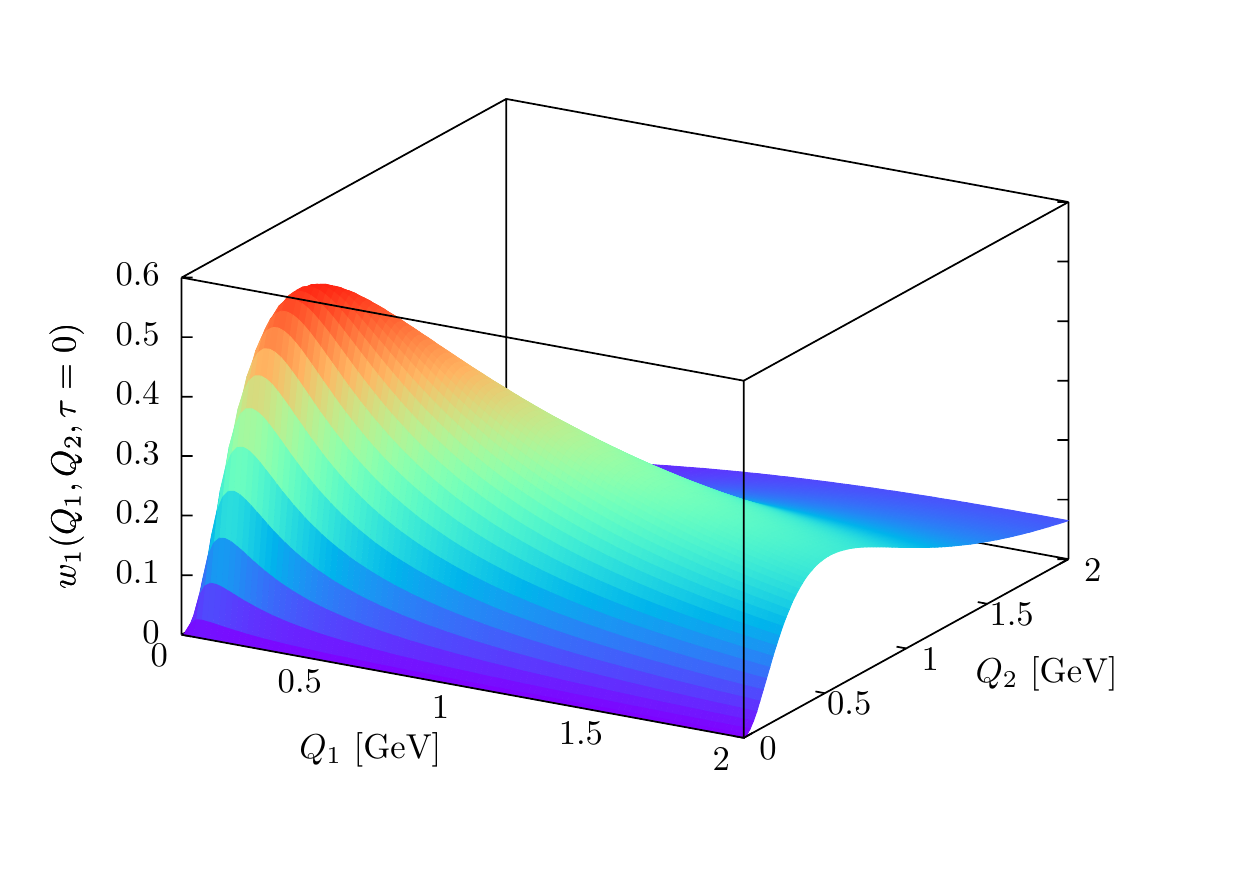}
\hspace*{1cm}\includegraphics[height=5cm,clip]{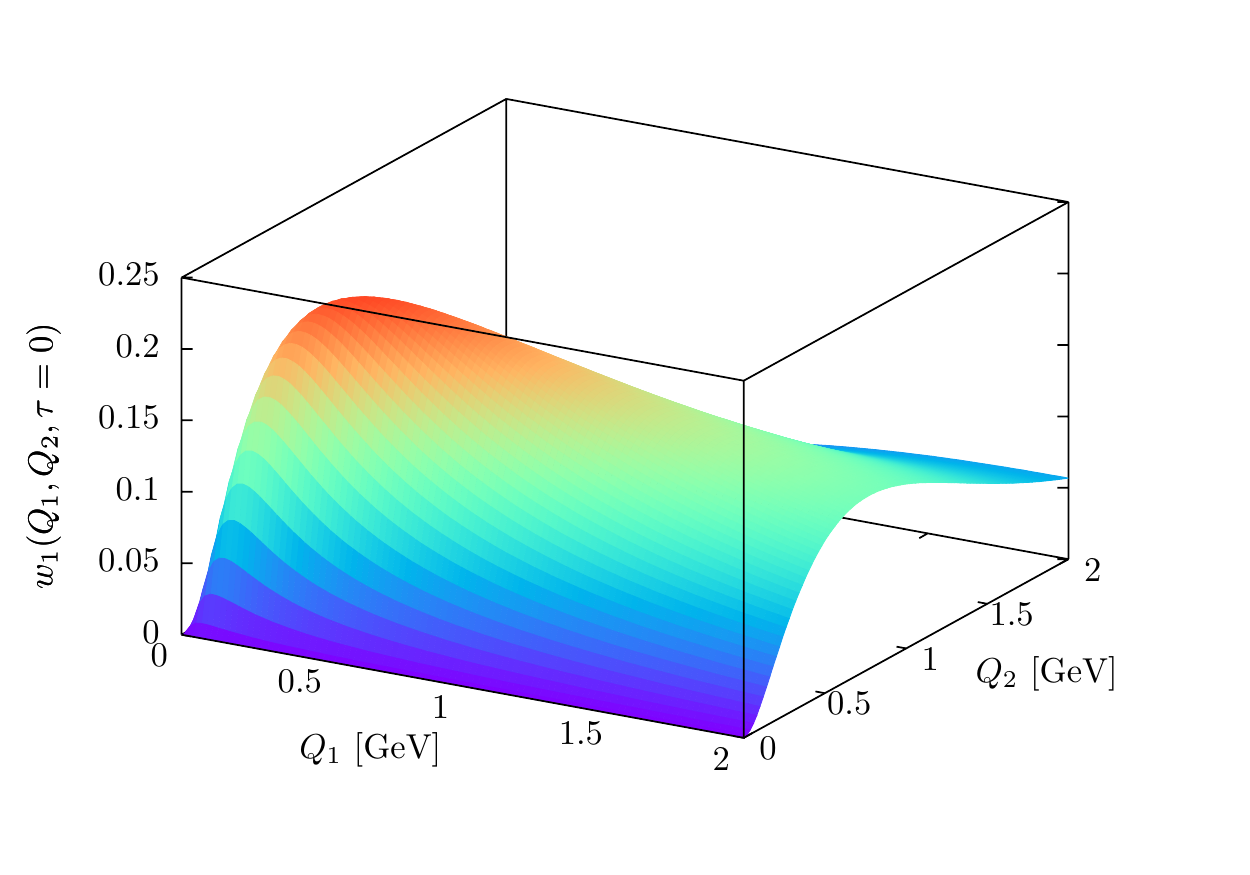}
\caption{Model-independent weight functions for the $\pi^0$:
  $w_1(Q_1,Q_2,\tau)$ (top left) and $w_2(Q_1,Q_2,\tau)$ (top right)
  as function of the Euclidean momenta $Q_1$ and $Q_2$ for $\theta =
  90^\circ~(\tau = 0)$.  Weight functions $w_1(Q_1,Q_2,\tau)$ for
  $\eta$ (bottom left) and $\eta^\prime$ (bottom right) for $\theta =
  90^\circ$. The weight functions $w_2(Q_1, Q_2, \tau)$ for $\eta$ and
  $\eta^\prime$ have a similar shape as for the pion, but the peaks
  are broader.}
\label{Fig:w1_w2}
\end{figure*}

In Fig.~\ref{Fig:w1_w2} we have plotted the weight functions
$w_1(Q_1,Q_2,\tau)$ and $w_2(Q_1,Q_2,\tau)$ for the light pseudoscalars
$\pi^0, \eta, \eta^\prime$ as function of $Q_1$ and $Q_2$ for $\theta
= 90^\circ~(\tau = 0)$. Three-dimensional plots for a selection of other
values of $\theta$ and one-dimensional plots as function of $\tau =
\cos\theta$ for some selected values of $Q_1$ and $Q_2$ can be found
in Ref.~\cite{Nyffeler_16}. Note that although the weight functions
rise very quickly to the maxima in the plots in Fig.~\ref{Fig:w1_w2},
the slopes along the two axis and along the diagonal $Q_1 = Q_2$
actually vanish for both functions~\cite{Nyffeler_16}. We
stress that these weight functions are completely independent of any
models for the form factors. 

We can immediately see from the weight functions for the pion that the
low-momentum region, $Q_{1,2} \leq 0.5~\mbox{GeV}$, is the most
important in the corresponding integrals~(\ref{amupi0_1}) and
(\ref{amupi0_2}) for $a_\mu^{\HLbLpi}$.  In $w_1(Q_1,Q_2,\tau)$ there
is a peak around $Q_1 \sim 0.15 - 0.19~\mbox{GeV}$, $Q_2 \sim 0.09 -
0.16~\mbox{GeV}$. The values of the maxima of the weight functions for
all light pseudoscalars and the locations of the maxima in the
$(Q_1,Q_2)$-plane for a selection of $\theta$-values have been
collected in Ref.~\cite{Nyffeler_16}.  For $w_1(Q_1, Q_2, \tau)$ and
$\theta \leq 150^\circ$, a ridge develops along the $Q_1$ direction
for $Q_2 \sim 0.18 - 0.26~\mbox{GeV}$ (maximum along the line $Q_1 =
2~\mbox{GeV}$). For larger values of $\theta$, the function looks more
and more symmetric in $Q_1$ and $Q_2$. This ridge leads for a constant
form factor to an ultraviolet divergence $(\alpha/\pi)^3 {\cal C}
\ln^2(\Lambda/m_\mu)$ for some momentum cutoff $\Lambda$ with ${\cal
  C} = 3 (N_c/(12\pi))^2 (m_\mu/F_\pi)^2 = 0.0248$, see
Refs.~\cite{KN_02,Knecht_et_al_PRL_02}.  Of course, realistic form
factors fall off for large momenta and $a_\mu^{\HLbLpione}$ will be
convergent.

The weight function $w_2(Q_1,Q_2,\tau)$ for the pion is about an order
of magnitude smaller than $w_1(Q_1,Q_2,\tau)$. There is no ridge,
since the function is symmetric under $Q_1 \leftrightarrow Q_2$. For
$\tau$ near $-1$, the peak is around $Q_1 = Q_2 \sim 0.14~\mbox{GeV}$
and broader than the one shown for $\theta = 90^\circ$. The location
of the peak moves to lower values $Q_1 = Q_2 \sim 0.04~\mbox{GeV}$
when $\tau$ grows towards $+1$.

The dependence of the weight functions on the pseudoscalar mass
through the propagators shifts the relevant momentum regions (peaks,
ridges) in the weight functions, and thus also in HLbL, to higher
momenta for $\eta$ compared to $\pi^0$ and even higher for
$\eta^\prime$, see Fig.~\ref{Fig:w1_w2}. It also leads to a
suppression in the absolute size of the weight functions due to the
larger masses in the propagators. This pattern is also visible in the
values for the contributions to $a_\mu$. For the bulk of the weight
functions (maxima, ridges) we observe the following approximate
relations (not necessarily at the same values of the momenta and
angles)
\be \label{w1_ratios} 
\left. w_1 \right|_{\eta} \approx \frac{1}{6} \left. w_1
\right|_{\pi^0},  \qquad 
\left. w_1 \right|_{\eta^\prime} \approx \frac{1}{2.5} \left. w_1
\right|_{\eta}.  
\ee 
Of course, the ratio of the weight functions is given by the ratio of
the propagators and is maximal at zero momenta and at that point equal
to the ratio of the squares of the masses, but at zero momenta the
weight functions themselves vanish. The combined effect is well
described by the relations in Eq.~(\ref{w1_ratios}). Furthermore, for
both $\eta$ and $\eta^\prime$, the weight function $w_2$ is about a
factor 20 smaller than $w_1$.

The peaks for the weight function $w_1(Q_1,Q_2,\tau)$ for $\eta$ and
$\eta^\prime$ are less steep, compared to $\pi^0$, and the ridge is
quite broad in the $Q_2$-direction, so that the weight function is
still sizeable, compared to the maximum, for $Q_2 =
2~\mbox{GeV}$. Furthermore, the ridge falls off only slowly in the
$Q_1$-direction. In particular for the $\eta^\prime$, the ridge for
$\theta \leq 75^\circ$ is almost as big as the maximum out to values
of $Q_1 = 2~\mbox{GeV}$. For $w_2(Q_1,Q_2,\tau)$ the peaks are broader
and larger momenta contribute, compared to the pion.

For the $\eta$-meson, the peak in the weight function
$w_1(Q_1,Q_2,\tau)$ is around $Q_1 \sim 0.32 - 0.37~\mbox{GeV}$, $Q_2
\sim 0.22 - 0.33~\mbox{GeV}$. The peak for the weight function
$w_2(Q_1,Q_2,\tau)$ is around $Q_1 = Q_2 \sim 0.14~\mbox{GeV}$ for
$\tau$ near $-1$ as for the pion. The location of the peak moves down
to $Q_1 = Q_2 \sim 0.06~\mbox{GeV}$ when $\tau$ is near $+1$.  For the
$\eta^\prime$, the peak in $w_1(Q_1,Q_2,\tau)$ now occurs for even
higher momenta, $Q_1 \sim 0.41 - 0.51~\mbox{GeV}$, $Q_2 \sim 0.31 -
0.43~\mbox{GeV}$. The locations of the peaks of $w_2(Q_1,Q_2,\tau)$ in
the $(Q_1,Q_2)$-plane follow a similar pattern as for the $\eta$
meson.

\section{Relevant momentum regions in $a_\mu^{\HLbLP}$}

In order to study the impact of different momentum regions on the
pseudoscalar-pole contribution, we need, at least for the integral
with the weight function $w_1(Q_1, Q_2, \tau)$ in
Eq.~(\ref{amupi0_1}), some knowledge on the form factor
$\FFP(-Q_1^2,-Q_2^2)$, since the integral diverges for a constant form
factor.\footnote{The integral with the weight function
  $w_2(Q_1,Q_2,\tau)$ in Eq.~(\ref{amupi0_2}) is finite and small for
  a constant form factor.}  For illustration we take for the pion two
simple models to perform the integrals: Lowest Meson Dominance with an
additional vector multiplet, LMD+V model, based on the Minimal
Hadronic Approximation to large-$N_c$ QCD matched to certain QCD
short-distance constraints from the operator product expansion (OPE),
see Refs.~\cite{KN_EPJC_01,KN_02} and references therein, and the
well-known Vector Meson Dominance (VMD) model.  Of course, in the end,
the models have to be replaced as much as possible by experimental
data on the double-virtual TFF or one can use a DR for the form factor
itself~\cite{pion_TFF_DR}.

The main difference of the models is a different behavior of the
double-virtual form factor for large and equal momenta. The LMD+V
model reproduces by construction the OPE, whereas the VMD TFF falls
off too fast:
\bea 
\FFa^{\rm LMD+V}(-Q^2,-Q^2) & \!\!\!\!\!\sim\!\!\!\!\! & \FFa^{\rm
  OPE}(-Q^2,-Q^2) \!\sim\! \frac{1}{Q^2}, \label{large_Q_LMD+V} \\   
\FFa^{\rm VMD}(-Q^2,-Q^2)   & \!\!\!\!\!\sim\!\!\!\!\! & \frac{1}{Q^4}, \,
\quad \mbox{for large~}Q^2. \label{large_Q_VMD}  
\eea 
Nevertheless for not too large momenta, $Q_1 = Q_2 = Q =
0.5~[0.75]~\mbox{GeV}$, the form factors $\FFa(-Q^2, -Q^2)$ in the two
models differ by only $3\%~[10\%]$,
see~\cite{Nyffeler_16}. Furthermore, both models give an equally good
description of the single-virtual TFF
$\FFa(-Q^2,0)$~\cite{KN_EPJC_01,KLOE2_impact}.

The LMD+V model was developed in Ref.~\cite{KN_EPJC_01} in the chiral
limit and assuming octet symmetry. This is certainly not a good
approximation for the more massive $\eta$ and $\eta^\prime$ mesons.
For the $\eta$ and $\eta^\prime$ meson we therefore simply take the
usual VMD model as already done in Refs.~\cite{KN_02,N_09}.

The two models yield the following results for the pole-contribution 
of the light pseudoscalars to HLbL (we only list here the central
values) 
\bea 
a_{\mu; {\rm LMD+V}}^{\HLbLpi} & = & 62.9 \times
10^{-11} \, , \label{amuLMD+V} \\ 
a_{\mu; {\rm VMD}}^{\HLbLpi} & = & 57.0 \times
10^{-11} \, , \label{amuVMD} \\
a_{\mu; {\rm VMD}}^{{\rm HLbL};\eta} & = & 14.5 \times
10^{-11}, \label{amueta} \\ 
a_{\mu; {\rm VMD}}^{{\rm HLbL};\eta^\prime} & = & 12.5 \times
10^{-11}.  \label{amuetaprime}   
\eea  

The results (\ref{amuLMD+V}) and (\ref{amuVMD}) for the pion-pole
contribution in the two models are in the ballpark of many other
estimates, but they also differ by $9.4\%$, relative to the LMD+V
result, due to the different high-energy behavior for the
double-virtual TFF in Eqs.~(\ref{large_Q_LMD+V}) and
(\ref{large_Q_VMD}) for $Q_{1,2} \geq 1~\mbox{GeV}$. In fact, the
pattern of the contributions to $a_\mu^{\HLbLpi}$ is to a large extent
determined by the model-independent weight functions
$w_{1,2}(Q_1,Q_2,\tau)$, which are concentrated below about
$0.5~\mbox{GeV}$, up to that ridge in $w_1$ along the $Q_1$
direction. As long as realistic form factor models for the
double-virtual case fall off at large momenta and do not differ too
much at low momenta, we expect similar results for the pion-pole
contribution at the level of $15\%$ which is in fact what is seen in
the literature~\cite{KN_02, JN_09, HLbL_talks}. Nevertheless, due to
the ridge-like structure in the weight function $w_1$, the high-energy
behavior of the form factors is relevant at the precision of $10\%$
one is aiming for.

For the $\eta$ and $\eta^\prime$, the results in Eqs.~(\ref{amueta})
and (\ref{amuetaprime}) are as expected from the discussion of the
relative size of the weight functions in Eq.~(\ref{w1_ratios}). The
result for $\eta$ is about a factor 4 smaller than for the pion with
VMD. The result for $\eta^\prime$ is only slighly smaller than for
$\eta$.  Note that the normalization of the TFF from $\Gamma({\rm P}
\to \gamma\gamma)$ and the momentum dependence due to different vector
meson masses for $\eta$ and $\eta^\prime$ also play a role for the
results in Eqs.~(\ref{amueta}) and (\ref{amuetaprime}).

\begin{figure*}[t!]
\centering
\includegraphics[height=6cm,clip]{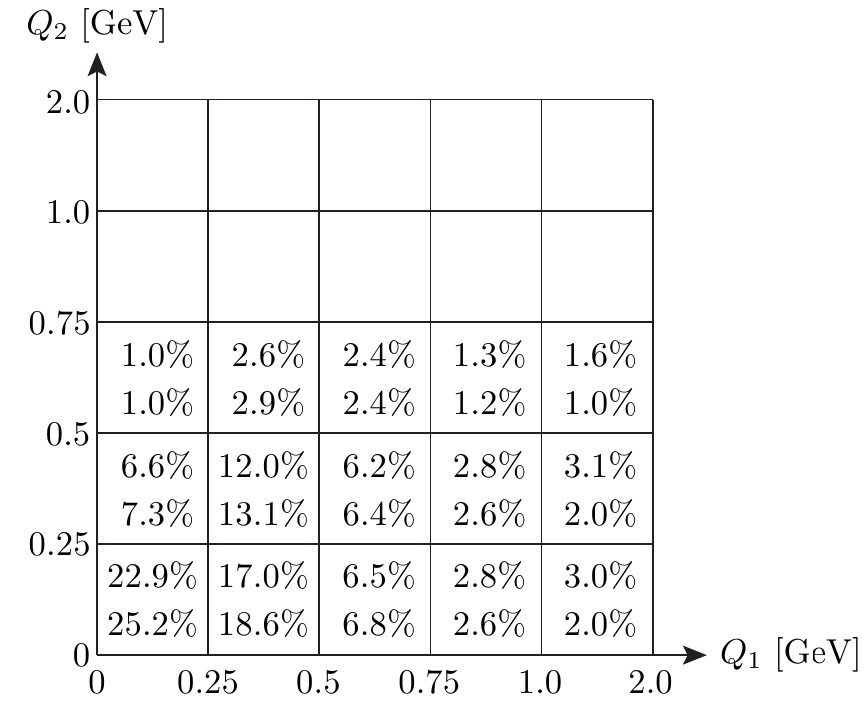}
\hspace*{1cm}\includegraphics[height=6cm,clip]{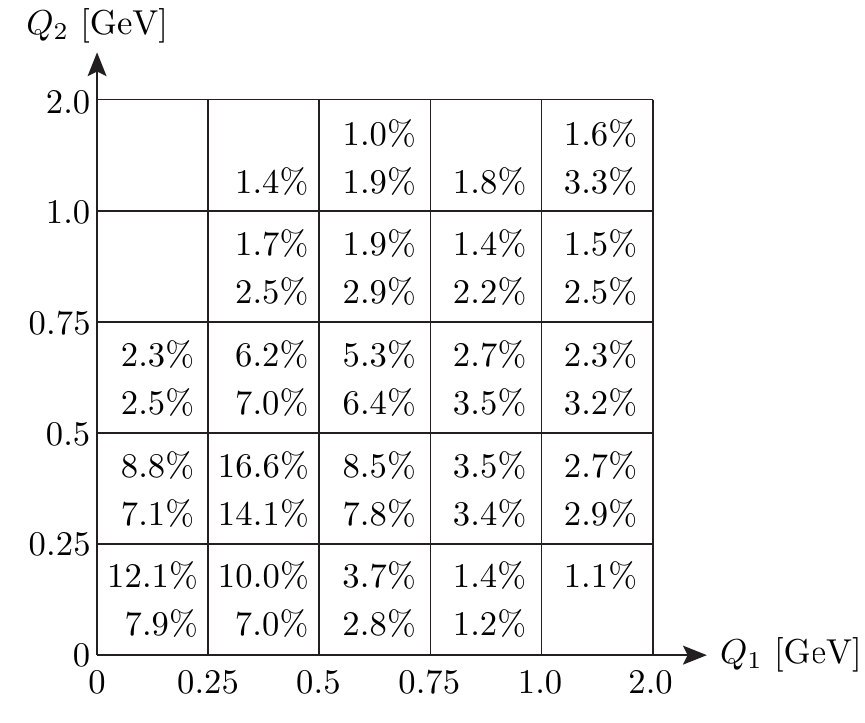}
\caption{Left panel: Relative contributions to the total
  $a_\mu^{\HLbLpi}$ from individual bins in the $(Q_1,Q_2)$-plane,
  integrated over all angles according to Eq.~(\ref{bins}). Note the
  larger size of the bins with $Q_{1,2} \geq 1~\mbox{GeV}$. Top line
  in each bin: LMD+V model, bottom line: VMD model. Contributions
  smaller than 1\% have not been displayed. For the LMD+V model there
  are further contributions bigger than 1\% along the $Q_1$-axis. For
  $2~\mbox{GeV} \leq Q_1 \leq 20~\mbox{GeV}$: $1.1\%$ in the bin $0
  \leq Q_2 \leq 0.25~\mbox{GeV}$ and $1.2\%$ for $0.25~\mbox{GeV} \leq
  Q_2 \leq 0.5~\mbox{GeV}$. Right panel: Relative contributions to the
  total of $a_\mu^{{\rm HLbL};\eta}$ and $a_\mu^{{\rm
      HLbL};\eta^\prime}$ with the VMD model from individual bins in
  the $(Q_1,Q_2)$-plane, integrated over all angles. Top line in each
  bin: $\eta$-meson, bottom line: $\eta^\prime$-meson.}
\label{Fig:PS_bins}
\end{figure*}

For a more detailed analysis, we integrate in Eqs.~(\ref{amupi0_1})
and (\ref{amupi0_2}) over individual momentum bins and all angles
$\theta$
\be \label{bins} 
\int_{Q_{1,{\rm min}}}^{Q_{1,{\rm max}}} dQ_1 \int_{Q_{2,{\rm
      min}}}^{Q_{2,{\rm max}}} dQ_2 \int_{-1}^{1} d\tau 
\ee 
and display the results, relative to the totals in
Eqs.~(\ref{amuLMD+V})-(\ref{amuetaprime}), in Fig.~\ref{Fig:PS_bins}.

Since the absolute size of the weight function $w_1(Q_1,Q_2,\tau)$ is
much larger than $w_2(Q_1,Q_2,\tau)$, the contribution from the integral
$a_\mu^{\HLbLpione}$ in Eq.~(\ref{amupi0_1}) dominates over
$a_\mu^{\HLbLpitwo}$ in Eq.~(\ref{amupi0_2}). Therefore the asymmetry
seen in the $(Q_1,Q_2)$-plane in Fig.~\ref{Fig:PS_bins}, with larger
contributions below the diagonal, reflects the ridge-like structure of
$w_1(Q_1,Q_2,\tau)$ in Fig.~\ref{Fig:w1_w2}. 

For the pion, the largest contribution comes from the lowest bin
$Q_{1,2} \leq 0.25~\mbox{GeV}$ since a large part of the peaks in the
weight functions (for different angles $\theta$) is contained in that
bin. More than half of the contribution comes from the four bins with
$Q_{1,2} \leq 0.5~\mbox{GeV}$.  In contrast, for the $\eta$ and
$\eta^\prime$, it is not the bin $Q_{1,2} \leq 0.25~\mbox{GeV}$ which
yields the largest contribution, since the maxima of the weight
functions are shifted to higher momenta, around
$0.3-0.5~\mbox{GeV}$. Furthermore, more bins up to $Q_2 =
2~\mbox{GeV}$ now contribute at least $1\%$ to the total. This is
different from the pattern seen for $\pi^0$.  The plots of the weight
functions for $\eta$ and $\eta^\prime$ in Fig.~\ref{Fig:w1_w2} show
that now the region $1.5 - 2.5~\mbox{GeV}$ also is important for the
evaluation of the $\eta$- and $\eta^\prime$-pole contributions. The
VMD model is, however, known to have a too fast fall-off at large
momenta, compared to the OPE.  Therefore the size of the contributions
$a_\mu^{{\rm HLbL};\eta}$ and $a_\mu^{{\rm HLbL};\eta^\prime}$ in
Eqs.~(\ref{amueta}) and (\ref{amuetaprime}) might be underestimated by
the VMD model, which could also affect the relative importance of the
higher momentum region in Fig.~\ref{Fig:PS_bins}.

Integrating both $Q_1$ and $Q_2$ from zero up to some upper momentum
cutoff $\Lambda$ and integrating over all angles $\theta$, one obtains
the results shown in Table~\ref{Tab:cutoffdependence}. This amounts to
summing up the individual bins shown in Fig.~\ref{Fig:PS_bins}.

\begin{table*} 
\centering 
\caption{Pseudoscalar-pole contribution $a_\mu^{\HLbLP} \times
  10^{11}, {\rm P} = \pi^0,\eta,\eta^\prime$ for different form
  factor models obtained with a momentum
  cutoff~$\Lambda$. In brackets, relative contribution of the total
  obtained with $\Lambda = 20~\mbox{GeV}$.}  
\label{Tab:cutoffdependence}
\renewcommand{\arraystretch}{1.1}
\begin{tabular}{r@{.}lr@{.}lr@{.}lr@{.}lr@{.}l}
\hline 
\multicolumn{2}{c}{$\Lambda$ [GeV]} & \multicolumn{2}{c}{$\pi^0$
  [LMD+V]}  & \multicolumn{2}{c}{$\pi^0$ [VMD]}  &
\multicolumn{2}{c}{$\eta$ [VMD]}  &  \multicolumn{2}{c}{$\eta^\prime$
  [VMD]}  \\ 
\hline 
0 & 25 & 14 & 4~(22.9\%) & 14 & 4~(25.2\%) &  1 & 8~(12.1\%) &  1 &
0~(7.9\%)  \\  
0 & 5  & 36 & 8~(58.5\%) & 36 & 6~(64.2\%) &  6 & 9~(47.5\%) &  4 &
5~(36.1\%) \\  
0 & 75 & 48 & 5~(77.1\%) & 47 & 7~(83.8\%) & 10 & 7~(73.4\%) &  7 &
8~(62.5\%) \\  
1 & 0  & 54 & 1~(86.0\%) & 52 & 6~(92.3\%) & 12 & 6~(86.6\%) &  9 &
9~(79.1\%) \\  
1 & 5  & 58 & 8~(93.4\%) & 55 & 8~(97.8\%) & 14 & 0~(96.1\%) & 11 &
7~(93.1\%) \\  
2 & 0  & 60 & 5~(96.2\%) & 56 & 5~(99.2\%) & 14 & 3~(98.6\%) & 12 &
2~(97.4\%) \\  
5 & 0  & 62 & 5~(99.4\%) & 56 & 9~(99.9\%) & 14 & 5~(100\%)  & 12 &
5~(99.9\%) \\  
20 & 0 & 62 & 9~(100\%)  & 57 & 0~(100\%)  & 14 & 5~(100\%)  & 12 &
5~(100\%) \\  
\hline 
\end{tabular}
\end{table*}

As one can see in Table~\ref{Tab:cutoffdependence}, for the pion more
than half of the final result stems from the region below $\Lambda =
0.5~\mbox{GeV}$ (59\% for LMD+V, 64\% for VMD) and the region below
$\Lambda = 1~\mbox{GeV}$ gives the bulk of the total result (86\% for
LMD+V, 92\% for VMD). The small difference between the form factor
models for small momenta $Q_{1,2} \leq 0.5~\mbox{GeV}$ is reflected in
the small absolute difference for $a_\mu^{\HLbLpi}$ in the two models
for $\Lambda \leq 0.75~\mbox{GeV}$. For instance, for $\Lambda =
0.75~\mbox{GeV}$, the difference is only $0.8 \times 10^{-11}$, i.e.\
$1.6\%$. The faster fall-off of the VMD model at larger momenta beyond
$1~\mbox{GeV}$, compared to the LMD+V model, leads to a smaller
contribution from that region to the total. Therefore we can see in
Fig.~\ref{Fig:PS_bins} that the main contributions in the VMD model,
relative to the total, are concentrated at lower momenta, compared to
the LMD+V model, in particular below $0.75~\mbox{GeV}$.

On the other hand, Table~\ref{Tab:cutoffdependence} shows that for
$\eta$ and $\eta^\prime$, the region below $\Lambda = 0.25~\mbox{GeV}$
only gives a small contribution to the total (12\% for $\eta$, 8\% for
$\eta^\prime$). Up to $\Lambda = 0.5~\mbox{GeV}$, we get about half
(one third) for $\eta$ ($\eta^\prime$) and the bulk of the results
comes from the region below $\Lambda = 1.5~\mbox{GeV}$, 96\% for the
$\eta$ and 93\% for the $\eta^\prime$-meson.

\section{Impact of transition form factor uncertainties on $a_\mu^{\HLbLP}$}

For the calculation of the pseudoscalar-pole contribution
$a_\mu^{\HLbLP}$ with ${\rm P} = \pi^0, \eta, \eta^\prime$ in
Eqs.~(\ref{amupi0_1}) and (\ref{amupi0_2}) the single-virtual form
factor $\FFP(-Q^2,0)$ and the double-virtual form factor $\FFP(-Q_1^2,
-Q_2^2)$, both in the spacelike region, enter.  We are interested here
in the impact of uncertainties of experimental measurements of these
form factors on the precision of $a_\mu^{\HLbLP}$. See Ref.~\cite{TFF}
for a brief overview of the various experimental processes where
information on the transition form factors can be obtained. In the
following, we will only quote the most relevant and most precise
experimental references. Ref.~\cite{Nyffeler_16} contains a detailed
analysis of the current experimental situation.

For the single-virtual form factor $\FFP(-Q^2, 0)$ the following
experimental information is available. The normalization of the form
factor can be obtained from the decay width $\Gamma({\rm P} \to
\gamma\gamma)$~\cite{PDG_2014, P_to_gamma_gamma}. Another important
experimental information is the slope of the form factor at the
origin. In the spacelike region, the slope and the form factor itself
have been measured in the process $e^+ e^- \to e^+ e^- \gamma^*
\gamma^* \to e^+ e^- {\rm P}$~\cite{TFF_spacelike}. The extraction of
the slope requires, however, a model dependent extrapolation from
rather large momenta $Q \sim 0.5~\mbox{GeV}$ to $Q^2 = 0$. The slope
and the TFF can also be obtained in the timelike region from the
single Dalitz-decay ${\rm P} \to \gamma^* \gamma \to \ell^+ \ell^-
\gamma$ with $\ell = e, \mu$~\cite{TFF_timelike}. Of course, for the
pion only the decay with an electron pair is possible. For the pion
the phase space is, however, rather small and the decay is not very
sensitive to form factor effects. The corresponding determinations of
the slope and the TFF are rather unprecise. The situation is much
better for $\eta$ and $\eta^\prime$. However, one then still needs to
perform an analytical continuation to obtain the form factor at
spacelike momenta. Recently a DR has been proposed in
Ref.~\cite{pion_TFF_DR} to determine the single- and double-virtual
form factor for the pion. So far, only the single-virtual form factor
has been evaluated in this dispersive framework with high precision at
low momenta $Q \leq 1~\mbox{GeV}$. For the $\eta$ and $\eta^\prime$, a
dispersive approach for the single- and double-virtual TFF has been
presented in Ref.~\cite{eta_etaprime_TFF_DR}.

For the single-virtual form factor we parametrize the measurement
errors in Eqs.~(\ref{amupi0_1}) and (\ref{amupi0_2}) as follows:
\be \label{delta1}
\FFP(-Q^2,0) \to \FFP(-Q^2,0) \, \left(1 \pm \delta_{1,{\rm P}}(Q)
  \right).  
\ee 
The momentum dependent errors $\delta_{1, {\rm P}}(Q)$ in different
bins are displayed in Table~\ref{Tab:delta1}, based on the analysis in
Ref.~\cite{Nyffeler_16}.  There are currently no experimental data for
the form factor available in the spacelike region in the lowest bin $0
\leq Q < 0.5~\mbox{GeV}$ for $\pi^0$ and $\eta$, except from
$\Gamma({\rm P} \to \gamma\gamma)$, the slope and timelike data for
the $\eta$-TFF. For the lowest bin we therefore assume an error, based
on ``extrapolating'' the current data sets and the data that will be
available soon from BESIII~\cite{BESIII_private} and maybe from
KLOE-2~\cite{KLOE2_impact} for the pion. If one uses the DR from
Ref.~\cite{pion_TFF_DR} below
$1~\mbox{GeV}$~\cite{pion_TFF_DR_private} and the current error on the
normalization from the decay width, one obtains (conservatively) the
uncertainties in the lowest two bins given in brackets in
Table~\ref{Tab:delta1}.
\begin{table}[h!] 
\centering 
\caption{Relative error $\delta_{1,{\rm P}}(Q)$ on the form factor
  $\FFP(-Q^2,0)$ for ${\rm P} = \pi^0, \eta, \eta^\prime$ in different momentum
  regions. The errors for $\delta_{1,\pi^0}(Q)$ and
  $\delta_{1,\eta}(Q)$ below $0.5~\mbox{GeV}$ are based on
  assumptions. In brackets for $\pi^0$ the uncertainties with a DR for
  the TFF.}    
\label{Tab:delta1}
\renewcommand{\arraystretch}{1.1}
\begin{tabular}{r@{$~Q~$}lr@{\%}lr@{\%}lr@{\%}l}
\hline 
\multicolumn{2}{c}{Region [GeV]} &
\multicolumn{2}{c}{$\delta_{1,\pi^0}(Q)$} &
\multicolumn{2}{c}{$\delta_{1,\eta}(Q)$}  &  
\multicolumn{2}{c}{$\delta_{1,\eta^\prime}(Q)$}  \\ 
\hline
$0 \leq $   & $ < 0.5$ & 5 & ~[2\%] & ~10 & &   6 & \\ 
$0.5 \leq $ & $ < 1$   & 7 & ~[4\%] & ~15 & & ~11 & \\ 
$1 \leq $   & $ < 2$   & 8 &        &   8 & &   7 & \\  
$2 \leq $   &          & 4 &        &   4 & &   4 & \\ 
\hline 
\end{tabular}
\end{table}

The second ingredient in Eqs.~(\ref{amupi0_1}) and (\ref{amupi0_2}) is
the double-virtual form factor $\FFP(-Q_1^2, -Q_2^2)$. Currently,
there are no direct experimental measurements available for this form
factor at spacelike momenta. For the pion, from the double Dalitz
decay $\pi^0 \to \gamma^* \gamma^* \to e^+ e^- e^+ e^-$, one can
obtain the double-virtual form factor $|\FFa(q_1^2, q_2^2)|$ at small
invariant momenta in the timelike region, but the results are
inconclusive~\cite{double_Dalitz}.  There is indirect information
available on the double-virtual TFF from the loop-induced decay ${\rm
  P} \to \ell^+ \ell^-~(\ell = e,\mu)$~\cite{PDG_2014,
  P_to_ll}. Without a form factor at the ${\rm
  P}-\gamma^*-\gamma^*$-vertex, the loop integral is ultraviolet
divergent. The relation between this decay and the pseudoscalar-pole
contribution $a_\mu^{\HLbLP}$ has been stressed in
Ref.~\cite{Knecht_et_al_PRL_02} and problems to explain both processes
simultaneously with the same model have been pointed
out~\cite{HLbL_PS_vs_lepton_pair_decay}.

In this situation, models have been used to describe the form factor
$\FFP(-Q_1^2, -Q_2^2)$ in the spacelike region and thus all current
evaluations of $a_\mu^{\HLbLP}$ are model dependent. Of course, it
would be preferable to replace these model assumptions as much as
possible by experimental data. In fact, it is planned to determine the
double-virtual form factor, at least for the pion, at BESIII for
momenta $0.5 \leq Q_{1,2} \leq 1.5~\mbox{GeV}$ and a first analysis is
already in progress~\cite{BESIII_private}, based on existing
data. 

\begin{figure*}[t]
\centering
\includegraphics[height=6cm]{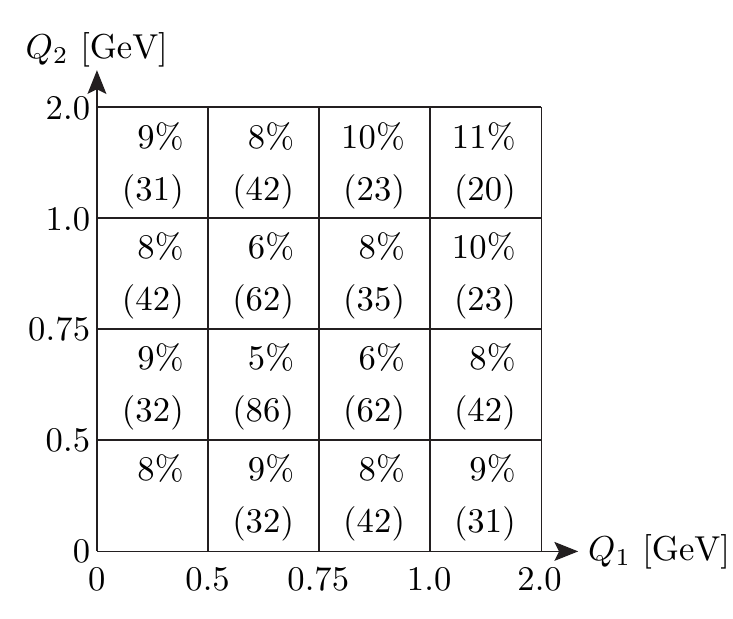}
\hspace*{1cm}\includegraphics[height=6cm]{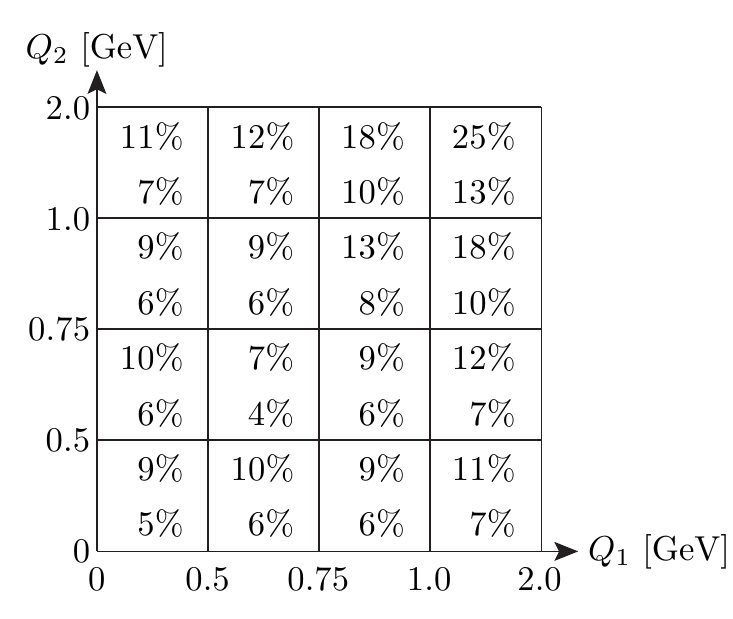}
\caption{Left panel: Assumed relative error $\delta_{2,
    \pi^0}(Q_1,Q_2)$ on the pion TFF $\FFa(-Q_1^2,-Q_2^2)$ in different
  momentum bins. Note the unequal bin sizes. In brackets the number of
  MC events $N_i$ in each bin according to the simulation with the
  LMD+V model for BESIII. For the lowest bin, $Q_{1,2} \leq
  0.5~\mbox{GeV}$, there are no events in the simulation due to the
  detector acceptance. In that bin, we assume as error the average of
  the three neighboring bins. For $Q_{1,2} \geq 2~\mbox{GeV}$, we take
  a constant error of 15\%. Right panel: Assumed relative error
  $\delta_{2, {\rm P}}(Q_1,Q_2)$ on the form factor $\FFP(-Q_1^2,-Q_2^2)$
  for $P = \eta, \eta^\prime$ in different momentum bins according to
  the MC simulations with the VMD form factor for $\eta$ (top line)
  and $\eta^\prime$ (bottom line). For $Q_{1,2} \geq 2~\mbox{GeV}$, we
  assume a constant error of 25\% for $\eta$ and 15\% for
  $\eta^\prime$.}
\label{Fig:FF_errors_bins}
\end{figure*}

In analogy to the single-virtual form factor, we parametrize potential
future measurement errors for the double-virtual form factor in
Eqs.~(\ref{amupi0_1}) and (\ref{amupi0_2}) in the following,
simplifying way: 
\bea 
\lefteqn{\FFP(-Q_1^2,-Q_2^2)} \nonumber \\
&& \to \FFP(-Q_1^2,-Q_2^2) \, \left(1 \pm 
  \delta_{2, {\rm P}}(Q_1,Q_2) \right), \label{delta2}
\eea
where the assumed momentum dependent errors $\delta_{2, {\rm
    P}}(Q_1,Q_2)$ in different bins are shown in
Fig.~\ref{Fig:FF_errors_bins}.
The error estimate for the form factor in each bin has been obtained
based on a Monte Carlo (MC) simulation~\cite{BESIII_private} for the
BESIII detector using the LMD+V model in the EKHARA event
generator~\cite{EKHARA} for the signal process $e^+ e^- \to e^+ e^-
\gamma^*\gamma^* \to e^+ e^- \pi^0$ and the VMD model for the
production of $\eta$ and $\eta^\prime$.  Since the number of events
$N_i$ in bin number $i$ is proportional to the cross-section
$\sigma_i$ (in that bin) and since for the calculation of the
cross-section the form factor enters squared, the statistical error on
the form factor measurement is given according to Poisson statistics
by $\sigma_i \sim \FFP^2 \Rightarrow \delta \FFP/\FFP = \sqrt{N_i}/(2
N_i)$.

In the lowest momentum bin $Q_{1,2} \leq 0.5~\mbox{GeV}$, there are no
events in the simulation, because of the acceptance of the
detector. When both $Q_{1,2}^2$ are small, both photons are almost
real and the scattered electrons and positrons escape detection along
the beam pipe. As a further assumption, we have therefore taken the
average of the uncertainties in the three neighboring bins as estimate
for the error in that lowest bin. This ``extrapolation'' from the
neighboring bins seems justified, since information along the two axis
is (or will soon be) available and the value at the origin is known
quite precisely from the decay width. Note that although the form
factor for spacelike momenta is rather smooth, it is far from being a
constant and some nontrivial extrapolation is needed.  For instance,
for the pion we get
$\FFa(-(0.5~\mbox{GeV})^2,-(0.5~\mbox{GeV})^2)/\FFa(0,0) \approx 0.5$,
for both the LMD+V and the VMD model~\cite{Nyffeler_16}.

The MC simulation~\cite{BESIII_private} corresponds to a data sample
of approximately half of the data collected at BESIII so far. The
simulation included only signal events. Based on a first preliminary
analysis of the BESIII data~\cite{BESIII_private} with strong cuts to
reduce the background from Bhabha events with additional photons, it
seems possible, at least for the pion, that the number of events and
the corresponding precision for $\FFa(-Q_1^2,-Q_2^2)$ shown in
Fig.~\ref{Fig:FF_errors_bins} could be achievable with the current
data set plus a few more years of data taking.  Of course, once
experimental data will be available, e.g.\ event rates in the
different momentum bins, there will still be the task to unfold the
data to reconstruct the form factor $\FFP(-Q_1^2, -Q_2^2)$ without
introducing too much model dependence.

Taking the LMD+V and VMD models for illustration, the assumed momentum
dependent errors from Table~\ref{Tab:delta1} and
Fig.~\ref{Fig:FF_errors_bins} impact the precision for the
pseudoscalar-pole contributions to HLbL as follows
\bea
a_{\mu; {\rm LMD+V}}^{\HLbLpi} & = & {62.9}^{+8.9}_{-8.2} \times
10^{-11} \quad \left(^{+14.1\%}_{-13.1\%}\right), \label{LMD+V_impact}
\\ 
a_{\mu; {\rm VMD}}^{\HLbLpi} & = & {57.0}^{+7.8}_{-7.3} \times 
10^{-11} \quad \left(^{+13.7\%}_{-12.7\%}\right), \label{VMD_impact}   
\\ 
a_{\mu; {\rm VMD}}^{{\rm HLbL};\eta} & = & 14.5^{+3.4}_{-3.0} \times
10^{-11} \quad \left(^{+23.4\%}_{-20.8\%}\right), \label{VMD_impact_eta}
\\ 
a_{\mu; {\rm VMD}}^{{\rm HLbL};\eta^\prime} & = & 12.5^{+1.9}_{-1.7}
\times 10^{-11} \quad
\left(^{+15.1\%}_{-13.9\%}\right). \label{VMD_impact_etaprime}  
\eea
While for the pion the absolute variations are different for the two
models, as are the central values, the relative uncertainty for both
models is around $14\%$. This will also be visible in the following
more detailed analysis. We therefore expect that using other form
factor models, and, eventually, using experimental data for the
single- and double-virtual form factors, will not substantially change
the following observations and conclusions.

More details have been collected in Table~\ref{Tab:amuHLbLP_impact},
which contains the results for the relative uncertainties from
Eqs.~(\ref{LMD+V_impact})-(\ref{VMD_impact_etaprime}) in the first
line. For the pion, the largest uncertainty of about $5\%$ comes from
the lowest bin $Q_{1,2} \leq 0.5~\mbox{GeV}$ in the $(Q_1,Q_2)$-plane
for $\delta_2$ in Fig.~\ref{Fig:FF_errors_bins} (fourth line in the
table).  Some improvement could be achieved, if the error in that
lowest bin and the neighboring bins could be reduced, to a total error
of about $12\%$, see the lines 8 and 9 in
Table~\ref{Tab:amuHLbLP_impact}. The second largest uncertainty of
$4.4\%$ for the pion stems from the lowest bin $Q < 0.5~\mbox{GeV}$ in
$\delta_1$ (second line in the table). Here the use of a dispersion
relation for the single-virtual form factor $\FFa(-Q^2, 0)$ for $Q <
1~\mbox{GeV}$ (see values in brackets in Table~\ref{Tab:delta1}) could
bring the total error of $14\%$ down to $11\%$, see the sixth line.

For the $\eta$ meson, the largest uncertainties of $7\%$ originate
from the region of $\delta_2$ above $0.5~\mbox{GeV}$ (5th line) and
from the lowest bin in $\delta_1$ (2nd line).  For the $\eta^\prime$,
the largest uncertainty of $5\%$ comes again from the region of
$\delta_2$ above $0.5~\mbox{GeV}$ (5th line). The second largest
uncertainty of $4.5\%$ comes from bins in $\delta_1$ above
$0.5~\mbox{GeV}$ (3rd line). For $\eta$ and $\eta^\prime$ the errors
go down to $20\%$ and $13\%$, if the uncertainty in the two lowest
bins in $\delta_1$ could be reduced, see line 7 in
Table~\ref{Tab:amuHLbLP_impact}. There is only a small reduction of
the uncertainty by one percentage point, if the errors in the lowest
few bins of $\delta_2$ could be reduced further, see lines 8 and 9.

\begin{table*}
\centering 
\caption{
  Impact of assumed measurement errors $\delta_{1, {\rm P}}(Q)$ and
  $\delta_{2, {\rm P}}(Q_1,Q_2)$ in the form factors $\FFP(-Q^2,0)$ and
  $\FFP(-Q_1^2, -Q_2^2)$ on the relative precision of the
  pseudoscalar-pole contributions (first line). Lines $2-5$ show
  the effects of uncertainties in different momentum regions below and above
  $0.5~\mbox{GeV}$. Lines $6-9$ show the impact of potential improvements
  of some of the assumed errors.}        
\label{Tab:amuHLbLP_impact}
\renewcommand{\arraystretch}{1.25}
\begin{tabular}{ccccl}
\hline 
$\frac{\delta a_{\mu; {\rm LMD+V}}^{\HLbLpi}}{a_{\mu; {\rm LMD+V}}^{\HLbLpi}}$ & 
$\frac{\delta a_{\mu; {\rm VMD}}^{\HLbLpi}}{a_{\mu; {\rm VMD}}^{\HLbLpi}}$  & 
$\frac{\delta a_{\mu; {\rm VMD}}^{{\rm HLbL};\eta}}{a_{\mu; {\rm
      VMD}}^{{\rm HLbL};\eta}}$ &  
$\frac{\delta a_{\mu; {\rm VMD}}^{{\rm HLbL};\eta^\prime}}{a_{\mu;
    {\rm VMD}}^{{\rm HLbL};\eta^\prime}}$ &  
\multicolumn{1}{c}{Comment} \\  
\hline 
${}^{+14.1\%}_{-13.1\%}$ & ${}^{+13.7\%}_{-12.7\%}$ & 
${}^{+23.4\%}_{-20.8\%}$ & ${}^{+15.1\%}_{-13.9\%}$ & 
Given $\delta_1,\delta_2$ \\
\hline
${}^{+4.3\%}_{-4.2\%}$ & ${}^{+4.4\%}_{-4.3\%}$ & 
${}^{+6.9\%}_{-6.8\%}$ & ${}^{+3.4\%}_{-3.3\%}$ &  
Bin $Q < 0.5~\mbox{GeV}$ in $\delta_1$ as given, rest: $\delta_{1,2} = 0$ \\ 
\hline 
${}^{+1.1\%}_{-1.0\%}$ & ${}^{+1.0\%}_{-0.9\%}$ & 
${}^{+4.4\%}_{-4.3\%}$ & ${}^{+4.5\%}_{-4.4\%}$ &  
Bins $Q \geq 0.5~\mbox{GeV}$ in $\delta_1$ as given, rest:
$\delta_{1,2} = 0$ \\ 
\hline 
${}^{+4.5\%}_{-4.4\%}$ & ${}^{+4.9\%}_{-4.8\%}$ & 
${}^{+4.0\%}_{-4.0\%}$ & ${}^{+1.7\%}_{-1.7\%}$ &  
Bin $Q_{1,2} < 0.5~\mbox{GeV}$ in $\delta_2$ as given, rest:
$\delta_{1,2} = 0$ \\ 
\hline 
${}^{+3.9\%}_{-3.8\%}$ & ${}^{+3.2\%}_{-3.1\%}$ & 
${}^{+7.0\%}_{-6.8\%}$ & ${}^{+5.1\%}_{-5.0\%} $ &  
Bins $Q_{1,2} \geq 0.5~\mbox{GeV}$ in $\delta_2$ as given, rest:
$\delta_{1,2} = 0$ \\ 
\hline 
${}^{+10.9\%}_{-10.5\%}$ & ${}^{+10.6\%}_{-10.1\%}$ &  
$-$ & $-$ & 
$\pi^0$: given $\delta_1,\delta_2$, but two lowest bins in
$\delta_{1,\pi^0}$ from DR: $2\%, 4\%$ \\ 
\hline 
$-$ & $-$ & 
${}^{+20.4\%}_{-18.5\%}$ & ${}^{+13.4\%}_{-12.5\%}$ &  
$\eta, \eta^\prime$: given $\delta_1,\delta_2$, but two lowest bins in 
$\delta_{1,\eta}$: $8\%, 10\%$ and $\delta_{1,\eta^\prime}$: $5\%, 8\%$  \\ 
\hline  
${}^{+12.4\%}_{-11.6\%}$ & ${}^{+11.8\%}_{-11.0\%}$ & 
${}^{+22.4\%}_{-20.0\%}$ & ${}^{+14.8\%}_{-13.6\%}$ & 
For $\pi^0, \eta, \eta^\prime$: given $\delta_1,\delta_2$, but lowest
bin in $\delta_2$: $8\%, 9\%, 5\% \to 5\%, 7\%, 4\%$ \\
\hline 
${}^{+12.0\%}_{-11.2\%}$ & ${}^{+11.4\%}_{-10.6\%}$ & 
${}^{+21.9\%}_{-19.6\%}$ & ${}^{+14.4\%}_{-13.4\%} $ & 
In addition,  bins in $\delta_2$ close to lowest bin: 
$9\%, 10\%, 6\% \to 5\%, 7\%, 4\%$ \\ 
\hline   
\end{tabular}
\end{table*}

\section{Conclusions} 

The three-dimensional integral representation for the
pseudoscalar-pole contribution $a_\mu^{\HLbLP}$ with ${\rm P} = \pi^0,
\eta, \eta^\prime$, from Ref.~\cite{JN_09} allows one to separate the
generic kinematics, described by model-independent weight functions
$w_{1,2}(Q_1, Q_2, \tau)$, from the double-virtual transition form
factors $\FFP(-Q_1^2, -Q_2^2)$. From the weight functions one deduces
that the relevant momentum regions are below $1~\mbox{GeV}$ for
$\pi^0$ and below about $1.5~\mbox{GeV}$ for $\eta$ and $\eta^\prime$.

If the assumed measurement errors $\delta_{1, {\rm P}}(Q)$ and
$\delta_{2, {\rm P}}(Q_1,Q_2)$ on the single- and double-virtual TFF
can be achieved in the coming years (in particular by measurements of
the double-virtual form factors at BESIII), one could obtain the
following, largely data driven, uncertainties for the
pseudoscalar-pole contributions to HLbL:
\bea 
\delta a_\mu^{\HLbLpi} / a_\mu^{\HLbLpi} & = & 14\% \quad
[11\%],  \\   
\delta a_\mu^{{\rm HLbL};\eta} / a_\mu^{{\rm HLbL};\eta} & = &
23\%,  \\   
\delta a_\mu^{{\rm HLbL};\eta^\prime} / a_\mu^{{\rm HLbL};\eta^\prime}
& = & 15\%.   
\eea 
The result in bracket for the pion uses the DR~\cite{pion_TFF_DR} for
the single-virtual TFF $\FFa(-Q^2, 0)$. Compared to the range of
estimates in the literature given in Eqs.~(\ref{range_HLbLpi0}) and
(\ref{range_HLbLP}) this would definitely be some progress. More work
is needed, however, to reach a precision of $10\%$ for all three
contributions. Experimental data on the double-virtual form factors in
the region $0 - 1.5~\mbox{GeV}$, e.g.\ from KLOE-2 or Belle 2, would
be very helpful in this respect.  A more detailed discussion can be
found in~\cite{Nyffeler_16}.

\begin{acknowledgement}
  I would like to thank the organizers for creating such a stimulating
  and nice atmosphere. I am grateful to Achim Denig, Christoph Redmer
  and Pascal Wasser for providing me with preliminary results of MC
  simulations for transition form factor measurements at BESIII and
  to Martin Hoferichter and Bastian Kubis for sharing information
  about the precision of the DR approach to the pion TFF. I thank the
  Heinrich Greinacher Foundation, University of Bern, Switzerland, for
  financial support. This work was supported by Deutsche
  Forschungsgemeinschaft (DFG) through the Collaborative Research
  Center ``The Low-Energy Frontier of the Standard Model'' (SFB 1044).
\end{acknowledgement}

% BibTeX or Biber users please use (the style is already called in the
% class, ensure that the "woc.bst" style is in your local directory) 
% \bibliography{name or your bibliography database}
%
% Non-BibTeX users please use

\end{document}